\documentclass[11pt]{article}

 \newcommand{\be}{\begin{equation}}
 \newcommand{\ee}{\end{equation}}
 \newcommand{\ba}{\begin{eqnarray}}
 \newcommand{\ea}{\end{eqnarray}}
 \newcommand{\bl}{\begin{equation}\begin{array}{ll}}
 \newcommand{\el}{\end{array}\end{equation}}
 \newcommand{\bll}{\begin{equation}\begin{array}{lll}}
 \newcommand{\bdm}{\begin{displaymath}}
 \newcommand{\edm}{\end{displaymath}}
 \def\p{\partial}
 \def\f{\varphi}
 \def\ve{\varepsilon}
 
\def\lim{\rightarrow}
\def\half{\frac{1}{2}}

\def\be{\begin{equation}}
\def\ee{\end{equation}}
\def\bea{\begin{eqnarray}}
\def\eea{\end{eqnarray}}

\def\dif{\partial}

\begin{document}
\raggedbottom

\title{Integrable Models of 1+1 Dimensional Dilaton \\
Gravity Coupled  to Scalar Matter}

\author{A.T.~Filippov\thanks{Alexandre.Filippov@jinr.ru}~ \\
{\small \it {$^+$ Joint Institute for Nuclear Research, Dubna,
Moscow Region RU-141980} }}

\maketitle

\begin{abstract}

A class of explicitly integrable models of 1+1 dimensional dilaton gravity
coupled to scalar fields is described in some detail. The equations
of motion of these models reduce to systems of the Liouville equations
endowed with energy and momentum constraints. The general solution
of the equations and constraints in terms of chiral moduli fields
is explicitly constructed and some extensions of the basic integrable
model are briefly discussed. These models may be related to high
dimensional supergravity theories but here they are mostly considered
independently of such interpretations. A brief review of other integrable
models of two - dimensional dilaton gravity is also given.

\end{abstract}

\section{Introduction}

It is well known that 1+1 dimensional dilaton
gravity coupled to scalar matter fields is a reliable
model for some aspects of high dimensional black holes,
cosmological models and branes.
The connection between high and low dimensions has
been demonstrated in different contexts of gravity and string
theory and in some cases allowed to find general solution
or some special classes of solutions in high dimensional
theories\footnote{For a more detailed discussion of this connection
and references see e.g. \cite{VA}, \cite{Kummer} and next section }.
For example, spherically symmetric gravity coupled to Abelian gauge
fields and massless scalar matter fields exactly reduces to a 1+1
dimensional dilaton gravity and can be explicitly solved if the
scalar fields are constants independent of coordinates.
Such solutions may describe some interesting physical objects --
spherical static black holes, simplest cosmologies, etc.
However, when the scalar matter fields, which presumably play a
significant cosmological role, are not constant, few exact analytical
solutions of high dimensional theories are known.
Correspondingly, the two - dimensional models of
dilaton gravity nontrivially coupled to scalar matter are usually not
integrable.

To obtain integrable models of this sort one has to make
some serious approximations, in other words, to deform the
original two - dimensional model obtained by a direct dimensional
reductions of realistic higher dimensional theories\footnote{
Note that some important four - dimensional space - times having
symmetries defined by two commuting Killing vectors may also be described
by two - dimensional dilaton gravity. For example, the cylindrical
gravitational waves are described by the 1+1 dimensional dilaton gravity coupled
to one scalar field. The stationary axially symmetric pure gravity may be described
by a 0+2 dimensional dilaton gravity coupled to one scalar field (this may be
related to the previous cylindrical case by the analytic continuation of one space
variable to imaginary values). Similar but more general dilaton gravity models
were also obtained in string theory. Some of them may be solved by using
modern mathematical methods developed in the soliton theory.
}.
Nevertheless, these models may qualitatively
describe some special physically interesting solutions of higher
dimensional gravity or supergravity theories related to
the low energy limit of superstring theories.

When preparing and writing this report I often recalled very
stimulating talks with Misha Saveliev about the Liouville and other
integrable theories. I cannot express my gratitude to him in
this world but I wish to mention that he strongly influenced
my interest and my approach to matters discussed below.

\section{Some integrable models of 1+1 dimensional dilaton
gravity coupled to scalar matter}

The effective Lagrangian of the 1+1 dimensional dilaton gravity
coupled to scalar fields $\psi_n$ that may be
obtained by dimensional reductions of a higher dimensional spherically
symmetric (super)gravity can usually be (locally) transformed to
the following form\footnote{
For a detailed motivation and specific examples see \cite{VA},
where references to other related papers may be found.
Due to space limitations, only absolutely necessary references are
given here.}
\be
{\cal L}^{(2)} = \sqrt{-g}\left[ \f R(g) + V(\f,\psi) +
\sum_n Z_n(\f, \psi)\, g^{ij} \, {\dif}_i \psi_n \, {\dif}_j \psi_n \right] \, .
\label{eq:7}
\ee
Here $g_{ij}(x^0,x^1)$ is a generic 1+1  metric with signature (-1,1),
$g \equiv {\rm det}|g_{ij}|$ and $R$ is the Ricci curvature of the
two - dimensional space - time,
\be
ds^2=g_{ij}\, dx^i \, dx^j \, , \,\,\,\,\,\, (i,j = 0,1).
\label{eq:4}
\ee
The effective potentials $V$ and $Z_n$
depend on the dilaton $\f (x^0,x^1)$ and on $N-2$ scalar fields
$\psi_n(x^0,x^1)$ (note that $Z_n < 0$).
They may depend on other parameters characterizing the parent higher
dimensional theory (e.g. on charges introduced by solving the equations
for the Abelian fields). Here we mainly consider the simplest kinetic terms,
 with $Z_n(\f, \psi) = Z_n(\f)$, or
 even with constant $Z_n$ that are independent of the fields.
We also used in (\ref{eq:7}) a Weyl transformation
to exclude the gradient term for the dilaton.

To simplify derivations we will use the equations of motion in the
light - cone metric, $ds^2 = -4f(u, v) \, du \, dv$.
By first varying this Lagrangian in generic coordinates and then
going to the light - cone ones we obtain the equations of motion
 \be
 \p_u \p_v \f+f\, V(\f,\psi)=0, \label{F.15}
 \ee
  \be
  f \p_i ({{\p_i \f} / f }) \, = \sum Z_n (\p_i \psi_n)^2\, , \,\,\,\,\,\,\,\,\,
 (i=u,v) \ .
 \label{F.17}
 \ee
 \be
\p_v (Z_n \p_u \psi_n) +\p_u (Z_n \p_v \psi_n) + f V_{\psi_n}(\f,\psi)=
\sum_m Z_{m, \psi_n} \, \p_u \psi_m \, \p_v \psi_m \ ,
 \label{F.16}
 \ee
 \be
 \p_u\p_v\ln |f| + f V_{\f}(\f,\psi) = \sum Z_{n, \f} \,\p_u \psi_n \, \p_v\psi_n \ ,
 \label{F.18}
 \ee
 where $V_{\f}= \p_{\f} V$, $V_{\psi_n} = \p_{\psi_n} V$,
 $Z_{n, \f}= \p_{\f} Z_n$, and $Z_{m, \, \psi_n} =\p_{\psi_n} Z_m$.
These equations  are not independent. Actually,
(\ref{F.18}) follows from (\ref{F.15}) $-$  (\ref{F.16}). Alternatively,
if  (\ref{F.15}), (\ref{F.17}), (\ref{F.18}) are satisfied,
one of the equations (\ref{F.16})
is also satisfied.

Consider the solution with constant scalar field
$\psi \equiv \psi_0$ (the `scalar vacuum' solution). This solution
exists if $V_{\psi}(\varphi,\psi_0) = 0$, see eq.(\ref{F.16}).
 The constraints (\ref{F.17}) can now be solved because their right-hand
sides are identically zero. It is a simple exercise to prove that there exist chiral
fields $a(u)$ and $b(v)$ such that $\varphi (u,v) \equiv \varphi (\tau)$
and $f(u,v) \equiv \varphi^{\prime} (\tau) \, a'(u) \, b'(v)$,
    where $\tau \equiv a(u) + b(v)$
(the primes denote derivatives with respect to the corresponding
argument). Using this result it is easy to prove that (\ref{F.15})
has the integral $ \varphi^{\prime} + N(\varphi) = M $,
where $N(\varphi)$ is defined by the equation $N^{\prime}(\f) = V(\f , \psi_0)$
and $M$ is the constant (integral) of motion.
The horizon, defined as a zero
of the metric $h(\tau)\equiv M - N(\varphi)$, exists because the equation
$M = N(\varphi)$ has at least
one solution in some interval of values of $M$. These solutions are actually
one - dimensional (`automatically' dimensionally reduced) and can be interpreted
    as black holes (Schwarzschild, Reissner Nordstr{\o}m, etc.)
or as cosmological models. Actually, the scalar vacuum theory is equivalent to the
pure dilaton gravity\footnote{
If $V = V(\f)$ and $Z_n \equiv 0$
the theory (\ref{eq:7}) is called the pure dilaton gravity.
It is integrable with arbitrary potential $V(\f)$.
},
which is a topological theory, and this explains
this drastic simplification of the equations of motion.
These facts are known for long time and were derived
by many authors using different approaches
(for references see e.g the recent
review \cite{Kummer}).

The properties of the general dilaton gravity theories (\ref{eq:7})
are much more complex. They do not reduce to one - dimensional models
and are in general not integrable.
If $Z_n \neq 0$ the theory may be integrable only with
very special potentials $V(\f,\psi)$ and
$Z_n(\f, \psi)$. Roughly speaking, the two - dimensional
models may be integrable if either the potentials $Z_n$
can be transformed to constants or the potential
$V$ is zero. In the first case, $V$
should have very special form that is described below.
In the second case,
for the theory to be explicitly analytically
integrable\footnote{We call the theory explicitly integrable
if one can explicitly write its general solution analytically,
in terms of a sufficient number of arbitrary functions,
in our case, of chiral free fields
    and of their derivatives. For further use in physics
problems we even need to have simple enough expressions for
the general solution.}
the potentials $Z_n$ must be very special functions.

In search for integrable 1+1 dimensional models
it may be useful to also consider `mixed' cases. Roughly speaking,
this means the following. Suppose we have three different species of
scalar matter fields -- $\psi_n$, $\chi_k$, and $\sigma_s$.
The characteristic properties of these fields are: the potential $V$
depends on $\f$ and $\psi_n$; $Z_n$ and $Z_k$ are constants, say,
$Z_n = Z_k =-1$; $Z_s = Z_s(\f)$ do not depend on $\sigma$
and $\psi$.
Thus the $Z$-part of the Lagrangian (\ref{eq:7}) is
\be
\sqrt{-g} \, g^{ij}\left[ \sum_n Z_n \, \p_i \psi_n \, \p_j \psi_n
+  \sum_k Z_k\, \p_i \chi_k \, \p_j \chi_k +
\sum_s Z_s(\f) \, \p_i \sigma_s \, \p_j \sigma_s \right]
\label{n1}
\ee
Looking at the equations (\ref{F.15}) - (\ref{F.18}) we see that $\chi_k$
are free massless fields (i.e. $\p_u \p_v \chi_k = 0$), the equations
(\ref{F.15}), (\ref{F.16}) for $\f$, and $\psi_n$ are
independent of the other fields, and, finally, the equations for $\sigma_s$
are linear. The right - hand sides of the constraints (\ref{F.17}),
\be
\sum_n Z_n \,  (\p_i \, \psi_n)^2
+  \sum_k Z_k \, (\p_i \, \chi_k)^2 +
\sum_s Z_s(\f) \, (\p_i \, \sigma_s)^2 , \,\,\,\,\,\, (i=u,v) ,
\label{n2}
\ee
depend on all three species of the matter fields,
so that the fields are not decoupled (this is natural because some
linear combinations of the constraints give the total energy and
momentum of the system that are constrained to zero). The most general
system of this sort may be explicitly integrable if its nonlinear sector
is explicitly integrable and the constraints can be explicitly solved
with arbitrary solutions for $\sigma$ and $\chi$ fields.
Unfortunately, in such a general setting the constraint cannot be solved.
As is shown in the next section, when we have only fields $f$, $\f$ and
$\psi$ (by taking constant $\sigma$ fields), each constraint involves
only functions of one variable ($u$ or $v$, resp.). However, this condition
cannot be satisfied for the term $Z_s(\f) \, (\p_i \, \sigma_s)^2 $ in
the integrable models discussed here. For this reason, mainly the
models with the matter fields $\psi$ and $\chi$ will be considered here.
Nevertheless, it is useful to keep $\sigma$-fields in mind for two reasons.
First, in some cases these terms may be taken into account as a perturbation.
The second reason is the following. In further reductions to
dimensions 0+1 or 1+0 the $\sigma$ fields may be exactly solved giving
additional terms to the potential $V$. The resulting theory may be integrable
or approximately integrable. We do not consider such possibilities here and
concentrate on the theories with the fields $\psi$ and $\chi$.

Let us first discuss the second `pure'  case (including $\chi$-fields
is trivial and we omit it). So, suppose that $V \equiv 0$,
$\psi \equiv \chi \equiv 0$, and $Z_s = Z(\f)$. Then, from the first
equation we have (up to the coordinate transformation
$u \mapsto a(u)$, $b \mapsto b(v)$): $\f = u+v \equiv r$
and the scalar fields satisfy the linear equation ($t \equiv u-v$)
\be
(\p_r^2 - \p_t^2)\sigma(r,t) +
{Z^{\prime}(r) \over Z(r)} \, \p_r \sigma(r,t) = 0 .
\label{n3}
\ee
When $Z(\f) = -\f$ this is the Euler -- Darboux
equation, for the general solution of which one can
write a rather complex integral representation
 (see e.g. \cite{Szekeres}). Moreover, the expression
for the metric is very difficult to analyze.
In general this does not allow us to find an explicit analytic expression
for the metric. Thus, such theories may be called integrable but
I would say that they are not explicitly integrable.
However, if $Z_n(\f)$ are solutions of the one - dimensional
Liouville equation, the potentials produced by such $Z_n(r)$,
are `non reflecting', the equations for $\sigma$ are explicitly
integrable and are simple functions of the chiral free massless
fields $a(u)$ and $b(v)$ (see \cite{FI}).
The metric is then explicitly expressed through
these chiral fields and thus the theory is explicitly
 (and elementary) integrable.
These solutions may describe plane waves of scalar matter coupled to
gravity or certain inhomogeneous cosmologies.

    The qualitative difference between the solutions of (\ref{n3}) for
    `non reflecting' potentials and the `realistic' ones (corresponding to
    $Z = -\f$) can be made more clear if we use the following beautiful,
    compact solutions of
    eq.(\ref{n3}) with $Z = -{\f}^{2\lambda + 1}$ ($\lambda > -\half$).
    The first solution was essentially found by S.D.Poisson,
    \be
    \sigma(r,t) = \int_0^{\pi} d\alpha \ q(t - r \cos \alpha) \sin^{2\lambda} \alpha \ ,
    \label{n3a}
    \ee
    the second one was derived later by E.W.Hobson\footnote{
    Known from 19-th century, these solutions and their generalizations
    may be found in the reprint of a still useful old book \cite{Bateman}.},
    \be
    \sigma(r,t) = \int_0^{\infty} d\alpha \ q(t \pm r\cosh \alpha) \sinh^{2\lambda} \alpha \ ,
    \label{n3b}
    \ee
    where $q(x)$ is an arbitrary `generating function'.
    Now, the potential corresponding to $Z = -{\f}^{2\lambda + 1}$ becomes
    `non reflecting' if $\lambda = l + \half$, with  integer $l$.
    Indeed, then (\ref{n3}) is equivalent to
    \be
    (\p_r^2 - \p_t^2)\bar{\sigma}(r,t) -
    {l(l+1) \over r^2} \bar{\sigma}(r,t) = 0 \ ,
    \label{n3c}
    \ee
    where $\bar{\sigma} \equiv r^{l+1} \sigma$. The potential $l(l+1)/r^2$ is a special case of
    the `non reflecting potentials' considered in \cite{FI}, with which the
     solution of this wave
    equation can be explicitly written as a linear combination of a free massless
    field $\chi = a(t+r) + b(t-r)$ and of its derivatives (with coefficients
    depending on $r$). One can easily prove that the solutions
    (\ref{n3a}) and (\ref{n3b}) reduce to such local fields when $\lambda = l+\half$
    with integer $l$. If $l$ is non integer the solution is a continual superposition
    of waves with different `velocities' $\sec \alpha$ or $\cosh^{-1} \alpha$.
    In this sense, the solution is generally nonlocal and, in fact, not simpler than
    one used in \cite{Szekeres}. It may be instructive to rewrite Eq.(\ref{n3a}),
     with half - integer $\lambda$, in the form
    \be
    \sigma(r,t) = {1 \over r} \int_{t-r}^{t+r} dx q(x) [(x-(t-r)) ((t+r)-x)]^l \ .
    \label{n3d}
    \ee
    If $l=0$ we find that
    $$
    \sigma(r,t) = {1 \over r} [Q(t+r) - Q(t-r)] \ ,
    $$
    where $Q^{\prime}(x) \equiv q(x)$. Similarly, one can find the local expressions
    of $\sigma$ in terms of massless free fields for any integer $l$.

    Originally, the equations
    (\ref{n3a}) and (\ref{n3b}) were applied to description of simple
    cylindrical waves (see e.g. \cite{Bateman}). More recently, it has been
    shown that they can be generalized to describe rather complex cylindrical
    gravitational waves (see e.g. \cite{Woodhouse} in which a generalization of these
    classical solution in the context of the Penrose twistor program is also discussed).
    In fact, the simplest cylindrical gravitational waves derived by A.Einstein
    and N.Rosen (see \cite{Rosen}) may be found by solving just the equation
    (\ref{n3}). To find the most general cylindrical gravitational waves one has to solve
    the following nonlinear equation
    \be
    \p_t (rM^{-1} \p_t M) - \p_r (rM^{-1} \p_r M) = 0 \ ,
    \label{n3e}
    \ee
    where $M$ is a $2 \times 2$ symmetric matrix depending on the cylindrical metric.
    This equation may be solved by a generalization of the Poisson formula
    \cite{Woodhouse} or by more general methods developed in the theory of solitons.

    In fact, the equation (\ref{n3e}) describing cylindrical gravitational waves is a
    special case of the field equations of $\sigma$-models coupled to gravity.
    This equation and its generalizations may be attempted to solve by the
    inverse scattering methods (ISM) of the soliton theory.
    Applications of these methods started with papers \cite{BZ} and \cite{Maison}
    which consider integrability of the stationary axial gravity. Further results and
    references to numerous papers dealing with this subject may be found in
    \cite{NKS} and \cite{Alekseev}. The methods for solving the integrable
    two - dimensional theories reducible to solving equations like (\ref{n3e}) are not
    elementary and, for this reason, are not included in our classification of
    the matter terms (\ref{n1}) that may lead to elementary solvable two - dimensional
    dilaton gravity. In our notation, the models that may be integrated by ISM
    have $V=0$, $Z_n = Z_k =0$ but, in general, more complex structure of the
    $\sigma$-terms:
    \be
    \sum_{\alpha, \beta} Z_{\alpha \beta}(\f; \sigma) \, \p_i \sigma_{\alpha} \,
    \p_j \sigma_{\beta}
    \label{n1}
    \ee
    In simple cases one may transform these terms to the diagonal form
    similar to (\ref{n1}) but with the potentials $Z$ depending on $\sigma$-fields.
    For example, the Lagrangian (2.8) of the two - dimensional dilaton gravity
    derived in \cite{BM} for the stationary, axially symmetric Einstein gravity
    can easily be transformed to the form of our `general' dilaton gravity
    (\ref{eq:7}) but this does not help to solve the equations of motion by
    elementary methods.
    On the other hand, we will see that there exist reasonably wide class of physically
    interesting,
    integrable two - dimensional theories with simple (constant!) $Z_n$ but rather complex
    potentials $V$.

The best studied examples of such integrable
models (belonging to the first `pure' class, with constant $Z_n$) are
Callan - Giddings - Harvey - Strominger (CGHS) model ($V = g_0$ )
and the Jackiw-Teitelboim (JT) model
($V = g_1 \f$). In CGHS model $R=0$ and in JT model $R=-g_1$.
A strong generalization of these models, in which the
two - dimensional curvature is not conformal constant
($V = g_{+}e^{g\f} + g_{-}e^{-g\f}$)
was proposed by the present author in \cite{A1}. Like CGHS and JT
models, this model belongs to the `mixed' case having free fields
$\chi_k$. It was called the 2-Liouville model because the fields
$\psi_{\pm} \equiv \ln f \pm \f$ satisfy two Liouville equations.
The Liouville
equations are, of course, explicitly solved in terms of two pairs
of the chiral free massless fields $a_{\pm}(u)$, $b_{\pm}(v)$.
By using the constraints (\ref{F.17})
one of the free scalar fields $\chi$ can be expressed in terms
of $a$, $b$ and of other $\chi$-fields.
So one may be inclined to decide that the two - Liouville
model is explicitly solved in \cite{A1}. However, there remained one `small'
problem -- the right - hand sides of the constraints are negative
definite while the left - hand sides may change sign. In \cite{A1}
I did not find an explicit analytic expression for the
chiral fields $a$ and $b$ that fully satisfy the constraints.
The explicit solution of the constraints (\ref{F.17})
was constructed only five years later when I have found a much more general
$N$-Liouville model \cite{A2}, \cite{VA}. Here I will show how this class
of models may be constructed and give its complete solution in terms
of chiral moduli fields.

\section{N-Liouville 1+1 dimensional model and its solution}
Let us suppose that the theory is
defined by the Lagrangian,
\be
{\cal L}^{(2)} = \sqrt{-g} \left[ \f R(g) + V(\f,\psi) +
g^{ij} \left( \sum_n Z_n \, \p_i \psi_n \, \p_j \psi_n
+  \sum_k Z_k\, \p_i \chi_k \, \p_j \chi_k \right) \right] ,
\label{n4}
\ee
with the following potentials:
\be
|f|V = \sum_{n=1}^N 2g_n e^{q_n} \, , \,\,\,\;\;\;\;\, Z_n = Z_k =-1 \ .
\label{eq:8}
\ee
Here $f$ is the light - cone metric, $ds^2 = -4f(u, v) \, du \, dv$, and
\be
q_n \equiv F + a_n \f + \sum_{m=3}^{N} \psi_m a_{mn} \equiv
\sum_{m=1}^{N} \psi_m a_{mn} \, , \label{eq:9}
\ee
where $\psi_1 + \psi_2 \equiv \ln{|f|} \equiv F$
($f \equiv \varepsilon e^F$, $\varepsilon = \pm 1$),
$\psi_1 - \psi_2 \equiv \f$ and thus
$a_{1n} = 1 + a_n$, $a_{2n} = 1 - a_n$.
By varying the Lagrangian (\ref{n4}) in $N-2$ scalar fields, dilaton, and in
$g_{ij}$ and then passing to the light - cone metric we find $N$
equations of motion for $N$ functions $\psi_n$,
\be
\epsilon_n \p_u \p_v \psi_n  =  \sum_{m=1}^{N} \varepsilon g_me^{q_m} a_{mn}
\, ; \,\,\,\,
\epsilon_1 = -1, \,\,\, \epsilon_n = +1, \,\, {\rm if} \,\, n \geq 2 \, ,
\label{eq:10}
\ee
as well as two constraints,
\be
C_i \equiv f \p_i (\p_i \f /f) + \sum_{n=3}^N (\p_i \psi_n)^2 =
 4D_i(i) ,\,\,\,\,\,  i = (u, v) , \,\, \label{eq:11}
\ee
where $4D_i(i) \equiv -\sum (\p_i \, \chi_k)^2$ is the contribution of the
free fields $\chi$ (see eq.(\ref{n2})) that depends on one variable because
$\chi_k = \alpha_k (u) + \beta_k(v)$ (here and below $k=1,...,K$).

 With arbitrary coefficients $a_{mn}$ these equations of motion are not integrable.
 However, as proposed in \cite{A2}, the equations (\ref{eq:10}) are
 integrable and the constraints (\ref{eq:11}) can be solved if the $N$-component
 vectors $v_n \equiv (a_{mn})$ are pseudo - orthogonal.
 Then the equations (\ref{eq:10}) reduce to $N$ independent, explicitly integrable
 Liouville equations for $q_n$,
\be
\p_u \p_v q_n - {\tilde{g}}_n e^{q_n} =0 , \label{eq:12}
\ee
where ${\tilde{g}}_n = \ve \lambda_n g_n$, $\lambda_n = \sum \epsilon_m
a_{mn}^2$, and $\ve \equiv f/|f|$ (note that the equations for $q_n$
depend on $\epsilon_n$ and $a_{mn}$ only implicitly, through the normalization factor
$\lambda_n$)\footnote{Here we suppose that $\lambda_n \neq 0$ and $g_n \neq 0$.
Otherwise the solution of the constraints should be modified in a fairly
 obvious way. We also denote $\gamma_n \equiv {\lambda_n}^{-1}$.}.

 The expression for the original fields in terms
 of the Liouville fields $q_n$
 may be found by using the orthogonality relations
 for $a_{mn}$ ($m \neq n$) and the definition of
 $\lambda_n \equiv \gamma_n^{-1}$ (for $m=n$)
 combined in the equation:
    \be
\sum_{l=1}^{N} \epsilon_l a_{lm} a_{ln} = \lambda_n \delta_{mn} \equiv
\gamma_n^{-1} \delta_{mn} \ .
\label{eq:12b}
    \ee
This equation may be written in the matrix form, if we define
the matrices $a = (a_{mn})$, $\gamma = (\gamma_m \delta_{mn})$ and
$\epsilon = (\epsilon_m \delta_{mn})$:
\be
a^T \epsilon a = \gamma^{-1} \ , \,\,\,\,\,\,\,\, a \gamma a^T = \epsilon .
\label{ad1}
\ee
It is important to note that in addition to these relation the matrix $a$
satisfies two conditions $a_{1n} = 1+a_n$, $a_{2n} = 1-a_n$.

Now, using (\ref{eq:12b}) or (\ref{ad1}) we can invert the
definition (\ref{eq:9}) and get
\be
\psi_m = \epsilon_m \sum_{n=1}^N a_{mn} \gamma_n q_n \ , \,\,\,\,\,
F = -2 \sum_{n=1}^N \gamma_n a_n q_n \ , \,\,\,\,\,
\f = -2 \sum_{n=1}^N \gamma_n q_n \ .
\label{eq:12aa}
\ee
 Writing the first equation for $n = 1,2$, expressing $a_{1n}$, $a_{2n}$
 in terms of $a_n$ and using linear independence of the functions $\psi_n$
 we may find very useful identities (`sum rules') for
 $\gamma_n , a_n$  and $a_{mn}$  for  $m \geq 3$. We have
\be
\psi_1 \equiv -\psi_1 \sum_{n=1}^N \gamma_n (1+a_n)^2 -
\psi_2 \sum_{n=1}^N \gamma_n (1-a_n^2) -
\sum_{m\geq 3}^N \psi_m \sum_{n=1}^N a_{mn} (1 + a_n) \gamma_n \ ,
\label{ad2}
\ee
\be
\psi_2  \equiv \,\,\,\, \psi_1 \sum_{n=1}^N \gamma_n (1-a_n^2)  +
\psi_2 \sum_{n=1}^N \gamma_n (1-a_n)^2 +
\sum_{m\geq 3}^N \psi_m \sum_{n=1}^N a_{mn} (1 - a_n) \gamma_n \ .
\label{ad3}
\ee
Thus we immediately find that the following identities should be satisfied
    \be
\sum_{n=1}^N \gamma_n =0 ; \,\, \sum_{n=1}^N \gamma_n a_n = -\half ; \,\,
\sum_{n=1}^N \gamma_n a_n^2 = 0 ; \,\, \sum_{n=1}^N a_{mn} \gamma_n =
\sum_{n=1}^N a_{mn} a_n \gamma_n = 0 , \,  m \geq 3 \, .
\label{eq:12c}
    \ee

Also, using the orthogonality relations (\ref{eq:12b})
it can be proven that one and only one norm $\gamma_n$ is negative.
We thus choose $\gamma_1 < 0$ while other $\gamma_n$ are
positive. In physically motivated models the parameters
$a_{mn}$, $\gamma_n$ and $g_n$ may satisfy some further
relations. For example, the signs of $\gamma_n$ and
$g_n$ may be correlated so that $g_n / \gamma_n < 0$.
However, such relations do not follow from the orthogonality
conditions and we ignore them in our discussion\footnote{
In fact, the coupling constants have nothing to do with
integrability and may be arbitrary numbers.}.

The most important fact is that the constraints can be explicitly solved.
First, we write the solutions of the Liouville equations
(\ref{eq:12}) in the form suggested by the conformal symmetry properties
of the Liouville equation \cite{Gervais},
\be
e^{-q_n /2} = a_n(u) b_n(v) - \half {\tilde{g}}_n \bar{a}_n(u) \bar{b}_n(v) \equiv X_n(u,v) \ ,
\label{eq:13}
\ee
where the chiral fields $a_n(u)$, $b_n(v)$, $\bar{a}_n(u)$
and $\bar{b}_n(v)$ satisfy the equations (do not mix $a_n(u)$ with $a_n$ used above):
\be
a_n(u) \bar{a}_n^{\prime}(u) - a_n^{\prime}(u) \bar{a}_n(u) = 1 , \,\,\,\,\,\,
b_n(v) \bar{b}_n^{\prime}(v) - b_n^{\prime}(v) \bar{b}_n(v) = 1 .
\label{eq:13a}
\ee
Using (\ref{eq:13a}) we can express $\bar{a}$ and $\bar{b}$
in terms of $a$ and $b$ and thus write $X_n$ as
\be
X_n(u,v)  = a_n(u) b_n(v) \biggl[ 1 - \half {\tilde{g}}_n
\int {du \over a_n^2(u)} \int {du \over b_n^2(v)} \biggr] \, .
\label{eq:13b}
\ee
It is not so straightforward but not very difficult to
rewrite the constraints (\ref{eq:11}) in the form:
\be
  C_u \equiv 4\sum_{n=1}^N \gamma_n {{{a_n}^{\prime \prime} (u)} \over {a_n(u)}}
  = 4 D_u(u)
\, , \,\,\,\,\,\,
 C_v \equiv 4\sum_{n=1}^N \gamma_n {{{b_n}^{\prime \prime} (v)} \over {b_n(v)}}
 = 4 D_v(v) \, .
\label{eq:14}
\ee
We first note that from (\ref{eq:9}) and (\ref{ad1}) it is easy to find the identities
\be
-\p_i F \p_i \f + \sum_{n=3}^N (\p_i \, \psi_n)^2 =
 \sum_{n=1}^N \gamma_n (\p_i \, q_n)^2 \ .
\label{n5}
\ee
It follows that
\be
C_i = \sum_{n=1}^N \gamma_n ((\p_i \, q_n)^2 - 2\p_i^2 q_n) =
4\sum_{n=1}^N \gamma_n {\p_i^2 X_n / X_n} \ .
\label{n6}
\ee
Now, from the definition of $X_n$ it is not difficult to see that
\be
{{\p_u^2 X_n} \over X_n} = {{a_n^{\prime \prime} (u)} \over {a_n(u)}} \ ,
\,\,\,\,\,\,\,\,\,\,\,\,\,
{{\p_v^2 X_n} \over X_n} = {{b_n^{\prime \prime} (v)} \over {b_n(v)}} \ .
\label{n7}
\ee

Using the relation $\sum \gamma_n = 0$
we can explicitly solve these constraints.
We show this for $C_u$ as the derivation for $C_v$ is similar.
Let us introduce temporary notation $a_n^{\prime}(u) / a_n(u) \equiv r_n(u)$,
use it in expression for $C_u$ and then shift all $r_n$ by an unknown
auxiliary function $R(u)$, i.e. define $\rho_n(u) \equiv r_n(u) + R(u)$.
Thus we may express $C_u$ in terms of $\rho_n(u)$ and $R(u)$:
\be
{1 \over 4} C_u = \sum_{n=1}^N \gamma_n {{{a_n}^{\prime \prime} (u)} \over {a_n(u)}} =
\sum_{n=1}^N \gamma_n (r_n^{\prime} + r_n^2) =
\sum_{n=1}^N \gamma_n [\rho_n^{\prime}(u) + \rho_n^2(u) -2 \rho_n(u) R(u)] \ ,
\label{n7a}
\ee
where we used the identity $\sum \gamma_n = 0$.
Now it is easy to see that the constraint $C_u = 4D_u$ will be solved if we take
\be
R(u) = \half \left[ \sum_{n=1}^N \gamma_n (\rho_n^{\prime}(u) + \rho_n^2(u))
- D_u(u) \right] \left[ \sum_{n=1}^N \gamma_n \rho_n(u)  \right]^{-1} \ ,
\label{n8}
\ee
where $\rho_n (u)$ are arbitrary functions. Thus, if we
choose $a_n^{\prime}(u) / a_n(u) = \rho_n - R(u)$, where $R$ is given
by (\ref{n8}), the constraint $C_u = 4D_u$ will be satisfied\footnote{
Note that the $\rho_n - R$ are not independent functions. In fact, the
chiral fields $a_n$ depend on $N-1$ arbitrary, independent functions
of $u$. This will be clear in a moment from a somewhat different
representation of the solution of the constraint.}.

Now we introduce new moduli fields
\be
\mu_m(u) \equiv \rho_m(u) - \half \left[ \sum_{n=1}^N \gamma_n  \rho_n^2(u)
- D_u(u) \right] \left[ \sum_{n=1}^N \gamma_n \rho_n(u)  \right]^{-1} \ ,
\label{n9}
\ee
Using this definition and the identity $\sum \gamma_n =0$,
it is easy to check that
\be
\sum \gamma_n \mu_n(u)  = \sum \gamma_n \rho_n(u) \ , \,\,\,\,\,\,\,\,
\sum \gamma_n \mu_n^2(u) = D_u(u) = -{1 \over 4} \sum (\alpha_k^{\prime}(u))^2
\label{n10}
\ee
(here and in what follows the limits of summation are omitted when the
summation extends over all possible values of $n=1,...,N$ and $k=1,...,K$).
Repeating this derivation for the second constraint $C_v$ and
defining the $v$-analogue, $\nu_n(v)$, of the moduli $\mu_n(u)$,
we finally write both constraints in terms of arbitrary chiral fields
$\mu_n(u)$, $\nu_n(v)$, $\alpha_k(u)$, $\beta_k(v)$:
\be
\sum \gamma_n \mu_n^2(u) + {1 \over 4} \sum (\alpha_k^{\prime}(u))^2  = 0 \ , \,\,\,\,\,\,\,\,
\sum \gamma_n \nu_n^2(v) + {1 \over 4} \sum (\beta_k^{\prime}(v))^2 =0 \ .
\label{n11}
\ee
These constraints are equivalent to the original ones, (\ref{eq:11}), if the
fields $f, \f, \psi, \chi$ are solutions of the equations of motion.

Returning to the original moduli fields $\rho_n$ and to their relation
to $a_n^{\prime}(u) / a_n(u) \equiv r_n(u)$ (and to their $v$-analogues),
one can see that the constraints are equivalent to the first order
differential equations for $a_n(u)$ and $b_n(v)$:
\be
{{a_n^{\prime} (u)} \over {a_n(u)}} = \mu_n (u) -
\half {{\sum \gamma_n {\mu_n}^{\prime}(u)} \over {\sum \gamma_n \mu_n(u)}} \, ,
\,\,\,\,\,\,\,
{{b_n^{\prime} (v)} \over {b_n(v)}} = \nu_n (v) -
\half {{\sum \gamma_n {\nu_n}^{\prime}(v)} \over {\sum \gamma_n \nu_n(v)}} \, ,
 \label{eq:15}
\ee
where $\mu_n(u)$ and $\nu_n(v)$ are arbitrary functions satisfying
the constraints (\ref{n11}). This is a small `miracle' allowing us
to explicitly solve the equations of motion (including the constraints)
in terms of a `sufficient' number of arbitrary functions.
The number of independent arbitrary chiral functions in r.h.s. of (\ref{eq:15})
is $2(N-1+K)$, where $N-2$ is the number of the scalar matter fields $\psi$,
and $K$ is the number of the free scalar matter fields $\chi$.
Together with the `gravitational' degrees of freedom $\f, F$
(or $\psi_1, \psi_2$) the theory without constraints would have
$2(N+K)$ chiral degrees of freedom. We will show in a moment that
the residual coordinate transformations allow us to further reduce
the number of independent arbitrary chiral functions to $2(N-2+K)$.

By integrating the first order differential equations
(\ref{eq:15}) for $a_n(u)$ and $b_n(v)$
we find the general solution of the $N$-Liouville dilaton gravity
in terms of the chiral moduli fields $\mu_n(u)$ and $\nu_n(v)$
satisfying the constraints (\ref{n11}):
\be
a_n(u) = {|\sum \gamma_m \mu_m|}^{-\half} \exp{\int du \, \mu_n(u)} , \,\,\,
b_n(v) = {|\sum \gamma_m \nu_m|}^{-\half} \exp{\int dv \, \nu_n(v)} .
\label{eq:15b}
\ee
Let us choose  arbitrary  chiral $\chi$-fields $\alpha_k(u)$ and $\beta_k(v)$.
Then the moduli fields $\mu_n(u)$ and $\nu_n(v)$ are not
independent due to the constraints (\ref{n11}).
In addition, we may use residual coordinate transformation
$u \rightarrow U(u)$ and $v \rightarrow V(v)$
to choose two gauge (coordinate) conditions. We will show in a moment
that writing
    \be
\mid \sum \gamma_n \mu_n(u)\mid \equiv U^{\prime}(u) \ , \,\,\,\,\,
\mid \sum \gamma_n \nu_n(v)\mid \equiv V^{\prime}(v)
\label{eq:15c}
    \ee
is indeed equivalent to choosing (U,V) as a new coordinate
system. With this aim, define new chiral fields,
\be
A_n \equiv \exp \int du \mu_n(u) \ , \,\,\,\,\,
B_n \equiv \exp \int dv \nu_n(v) \ ,
\label{eq:15d}
\ee
that we consider as the functions of the new coordinates $(U,V)$
(due to implicit transformations (\ref{eq:15c})).
Using (\ref{eq:15b}) and (\ref{eq:15c}) we may rewrite
(\ref{eq:13b}) in terms of $A_n(U)$, $B_n(V)$:
\be
Y_n(U,V) = A_n(U) B_n(V) \biggl[ 1 -
\half {\tilde{g}}_n \int {dU \over A_n^2(U)} \int {dV \over B_n^2(V)} \biggr] \, ,
\label{eq:15e}
\ee
where we have defined
\be
Y_n(U,V) \equiv  [U^{\prime}(u) V^{\prime}(v)]^{\half} X_n(u,v)
\label{eq:15ea}
\ee

Now, with the above definitions
we may derive the metric $f$,
\be
f (u,v) = U^{\prime}(u) V^{\prime}(v) \prod_{n=1}^{N} [Y_n(U,V)]^{4\gamma_n a_n} \equiv
\bar{f}(U,V) U^{\prime}(u) V^{\prime}(v) ,
\label{eq:15f}
\ee
the dilaton $\f$, and the scalar fields $\psi_m$ ($m\geq 3$):
\be
e^{\f} = \prod_{n=1}^{N} [Y_n(U,V)]^{4\gamma_n} , \,\,\,\,\,
e^{\psi_m} = \prod_{n=1}^{N} [Y_n(U,V)]^{-2a_{mn} \gamma_n} .
\label{eq:15g}
\ee
We see that $ds^2 = -4f(u,v)du dv  \equiv -4 \bar{f}(U,V) dU dV$
and thus everything is expressed in terms of the new
coordinates $(U,V)$.

The constraints (\ref{n11}) for the moduli parameters can
easily be solved by expressing one of the moduli in terms of others.
However, it may be more convenient and instructive to introduce new moduli
that make clear the topological nature of the moduli space and thus may allow
interesting topological classification of the solutions of the $N$-Liouville
theory. To simplify notation we define the new moduli for the case of zero
$\chi$-fields (the general case is not essentially different).  Thus, setting
$\chi \equiv 0$ we introduce the following unit $(N-1)$-vectors
(recall that $\gamma_1 <0$ and $\gamma_k > 0$ for $k \geq 2$):
\be
\hat{\xi}_k(u) \equiv {{\mu_k (u) \surd \gamma_k}\over {\mu_1 (u) \surd |\gamma_1}|} \ ,
\,\,\,\,\,\,\,\,\,
\hat{\eta}_k(v) \equiv {{\nu_k (v) \surd \gamma_k}\over {\nu_1 (v) \surd |\gamma_1}|} \ ,
\,\,\,\,\,\,\,\,\,  k=2,...,N \ .
\label{eq:15h}
\ee
These vectors, moving on the ($N-2$)-dimensional unit sphere
$S^{(N-2)}$, determine the solution up to a choice of the
coordinate system, which can be fixed by the above gauge
conditions (\ref{eq:15c}). They now look as follows:
\be
U^{\prime}(u) = |\gamma_1 \mu_1 (u)| (1 - \cos \theta_{\xi}(u))  , \,\,\,\,\,\,\,
V^{\prime}(v) = |\gamma_1 \nu_1 (v)| (1 - \cos \theta_{\eta}(v)) ,
\label{eq:15k}
\ee
\be
\cos \theta_{\xi}(u)  = \sum_{k=2}^N \hat{\gamma}_k \hat{\xi}_k(u)
\equiv \hat{\gamma} \hat{\xi} , \,\,\,\,\,\,\,
\cos \theta_{\eta}(v) = \sum_{k=2}^N \hat{\gamma}_k  \hat{\eta}_k(v) \equiv
\hat{\gamma}  \hat{\eta} ,
\label{eq:15l}
\ee
where $\hat{\gamma}$ is the constant unit $(N-1)$-vector,
$\hat{\gamma}_k = (\gamma_k / |\gamma_1|)^{\half}$.

The moduli fields $\hat{\xi}(u)$  and $\hat{\eta}(u)$ are very useful because
they give very simple representation of the most important solutions and,
in particular, visualize relations between solutions of different dimensions.
For example, if the vectors $\hat{\xi}(u)$  and $\hat{\eta}(v)$ are constant and
equal ($\hat{\xi} = \hat{\eta}$), they give a static or a cosmological solution.
The static solution has a horizon if $\hat{\xi} = \hat{\eta} = \hat{\gamma}$
(see \cite{VA}). Another interesting static solution, which is flat at infinity,
is given by the constant vectors satisfying the condition
$\hat{\xi} = \hat{\eta} = \hat{\delta}$, where
$\delta_k \equiv (\gamma_k a_k / |\gamma_1| a_1)^{\half}$ (it is not difficult
to check that this vector is indeed of unit length). Two - dimensional
solutions may be represented by pairs of curves in $S^{(N-2)}$ that we denote by
($\hat{\xi}(u)$, $\hat{\eta}(v)$). Those that asymptotically interpolate
between the above one - dimensional solutions are of special importance
and define additional structure on $S^{(N-2)}$, which may be used for
a physically motivated classification of the solutions of the $N$-Liouville
theory.

Especially interesting are the new solutions that, probably, may be localized
in space (`lumps' or soliton - like waves of scalar matter). They correspond
to constant but not equal vectors $\hat{\xi}$  and $\hat{\eta}$
(obviously, this is possible for $N \geq 3$). In the
coordinates $(U,V)$ ($U+V \equiv r$, $U-V \equiv t$) the solutions may be
written as
\be
Y_n(U,V) = C_n \cosh[R_n^{-1}((r - r_n) - v_n t)] ,
\label{n12}
\ee
where the parameters $C_n \ , R_n \ , r_n \ , v_n$ rationally depend
od the moduli and other parameters characterizing the Lagrangian and the
coordinate system (recall that this parameters are not independent, due to
the constraints and the gauge conditions).
Although the exponentials $e^q = Y_n^{-2}$ are localized in space, this is not
necessarily sufficient for the matter fields to be localized. However,
using freedom in choosing moduli and other parameters it may be possible
to write a truly localized solution. A detailed derivation and investigation
of these interesting solutions will be presented in a separate publication.

We thus have the general solution of the 1+1 dimensional dilaton gravity
coupled to any number of scalar fields.
It is explicitly expressed in terms
    of a sufficient number of arbitrary chiral fields\footnote{
    At first sight, the expressions for the fields $\bar{f}$, $\f$, $\psi_m$
    may look nonlocal. However, denoting the indefinite integrals in (\ref{eq:15e})
    by new chiral fields, we see that $Y_n$ are essentially local functions of
    new fields and of their derivatives.}
and thus we
may solve the Cauchy problem and study the evolution of cosmological or
black hole type solutions, etc. The representation of the general solution
in terms of the chiral fields $a_n(u)$ and $b_n(v)$ may give us a good starting
point in attempts to quantize our $N$-Liouville dilaton gravity.
Even more useful may be the chiral moduli fields $\mu_n(u)$ and $\nu_n(v)$
(or ${\hat{\xi}}_k(u)$ and ${\hat{\eta}}_k(v)$).
In terms of these moduli fields the dimensional reduction of the
solutions becomes very transparent and this may simplify the derivation and
 physical interpretation of the evolution of one - dimensional solutions
 and suggest new approaches to quantization  based on analogy with
the simple  1-dimensional case\footnote{
The one - dimensional models obtained by simple dimensional reduction
may easily be quantized. Our formulation is as close to the one - dimensional
theory as possible and this supports hopes to finally find a quantum version of
the $N$-Liouville theory. It should be emphasized that, up to now, the interesting
results obtained in quantum Liouville theory did not shed much light on the
$N$-Liouville theory. This may be explained by the fact that the main difficulty
and the main content of the $N$-Liouville, as distinct from the standard
Liouville theory, lies in the constraints.}.

\section{Discussion and outlook}
The explicitly analytically integrable models presented here
may be of interest for different applications.
Most obviously we may use them to construct first approximations to generally
non integrable theories. Realistic theories describing black holes and cosmologies
are usually not integrable.
However, explicit general solutions of the integrable approximations
may allow one to construct different sorts of perturbation theories.

For example, spherically symmetric static black holes non minimally coupled to
scalar fields are described by the integrable 0+1 dimensional $N$-Liouville
model. Similarly, spherically symmetric cosmological models may be described
by 1+0 dimensional $N$-Liouville theory.
However, the corresponding 1+1 theory is not integrable
because the scalar coupling potentials $Z_n$ are not constant
(actually, $Z_n \sim \f$).
To obtain approximate analytic solutions of the 1+1 theory one then
may try to approximate $Z_n$ by properly chosen constants.

This approach may be combined with the recently proposed analytic perturbation
theory allowing one to find solutions close to horizons
for the most general non integrable 0+1 dilaton gravity theories \cite{atfm}.
Near the horizons we can use the integrable 1+1 dimensional
dilaton gravity (with $Z_n = -1$) as a good approximation to a realistic
theory (with $Z_n$ depending on the dilaton $\f$).

In cosmological applications, the behaviour of the
1+1 dimensional solutions  near
the singularity at $\f =0$ is of great interest (see e.g. \cite{Venezia}).
Integrable 1+1 dimensional $N$-Liouville theories could give, at best,
a rough qualitative approximation of the exact solutions near
the singularity. A more quantitative approximation might be
obtained by first asymptotically solving the exact theory
in the vicinity of $\f =0$ and then sewing the asymptotic
solutions with those of the  integrable theory. To realize
such a program one needs a very simple and explicit analytic
solutions of the integrable theory. Our simple model having
the solutions represented in terms of the moduli $\hat{\xi}$
and $\hat{\eta}$ may give a good starting point for such a work.
Of course, before applications to realistic cosmologies become
possible, one should study in detail and completely classify
and interpret the behaviour of the 1+1 dimensional solutions
and their precise relation to the 1+0 dimensional reduction.

The reduction from dimension 1+1 both to dimension 1+0 (`cosmological')
and to dimension 0+1 (`static' or `black hole')
is especially transparent in the moduli representation
for the solutions of the 1+1 dimensional $N$-Liouville model.
However, as emphasized in \cite{VA}, the whole procedure of the
dimensional reduction should be reconsidered from a more general
point of view, taking into account more general dimensional reductions.
 A detailed consideration of generalized dimensional reductions
 will be given elsewhere.

\section{Appendix}
Here we sketch a simple approach to solving the pseudo orthogonality
conditions for $a_{mn}$ with additional restrictions $a_{1n} = 1 + a_n$,
$a_{2n} = 1 - a_n$. The equations one has to solve are
\be
a_i + a_j = \half \sum_{n=3}^N a_{ni} a_{nj} \ , \,\,\,\,\, i<j .
\label{a1}
\ee
These $N(N-1)/2$ equations are nonlinear but there exist a simple recursive
algorithm reducing their solution to solving linear equations. To find it one
has to choose which of the parameters should be regarded as unknown ones. Analyzing the
cases $N=3$ and $N=4$ one may find that the convenient choice of the division
of the parameters into unknown and arbitrarily fixed is the following.
Let us fix $a_{33}$ and $a_{mn}$ for $m>n,\ m\geq3$ and thus all other
parameters are unknown. First, consider the equations for $i,j =1,2,3$.
Denoting the r.h.s. of (\ref{a1}) by $A_k$ where $(ijk) = (123)_{\rm cyclic}$
we find
\be
2 a_i = -A_i +A_j +A_k \ , \,\,\,\,\,\, (ijk) = (123)_{\rm cyclic} \ .
\label{a2}
\ee
The next step is to consider the equations (\ref{a1}) for $i= 1,2,3$ and $j=4$.
These constitute three linear equations for three unknowns $a_4, a_{34}, a_{44}$.
They have a unique solution provided that the r.h.s.  and
the determinant $\Delta$ do not vanish.

Now the general construction should be clear. On the $j$-th step, where
$j=4,...,N$, we have $j-1$ equations for $j-1$ unknowns,
$a_j$ and $a_{nj}$ for $3 \leq n \leq j$:
\be
-2 a_j + \sum_{n=3}^j a_{ni} a_{nj} \ = \ 2a_i - \sum_{n=j+1}^N a_{ni} a_{nj} \ ,
 \,\,\,\,\, i=1,...,j-1 .
\label{a3}
\ee
The r.h.s. of these equations depend both on arbitrary and previously
found parameters. The nontrivial part of the procedure is that the
determinant, starting from $j=5$ depend on the previously found parameters
and thus the condition $\Delta \neq 0$ is difficult to control. For $j=4$
the determinant,
$$
\Delta = 2[(a_{33} - a_{31})(a_{43} - a_{42}) -
(a_{33} - a_{32}) (a_{43} - a_{41})] \ ,
$$
depends on the arbitrary parameters only.

Finally, let us write the parameters of the general model with $N=3$:
$$
2a_i = a_{3i} a_{3j} - a_{3j} a_{3k} + a_{3k} a_{3i} \ , \,\,\,\,
\gamma_i^{-1} = (a_{3i} - a_{3j})(a_{3i} - a_{3k}) \ , \,\,\,
(ijk) = (123)_{\rm cyclic} \ .
$$


\bigskip
\bigskip

 {\bf Acknowledgment:} The author appreciates financial support of
 the Dept. of Theoretical Physics of the University of Turin and
 INFN (Section of Turin), of CERN-TH, of the MPI and the W.Heisenberg
 Institute (Munich), where some results were obtained. Useful discussions with
 P.~Fr\'e, D.~Luest, D.~Maison, and G.~Veneziano are kindly acknowledged.
 Useful remarks of G.A.~Alekseev concerning Section~2 and references are
 also appreciated.
 Author is especially grateful to V.~de Alfaro for his support for many
 years and very fruitful collaboration; in particular, some of the results
 presented here were obtained in close collaboration with him.
 This report was completed while the author was visiting CERN-TH; kind
 hospitality and support of the members of TH is highly appreciated.
 This work was also partly supported by RFBR grant 03-01-00781-a.

\end{document}